\documentclass[twocolumn,showpacs,amsmath,amssymb,pra]{revtex4}

\usepackage{graphicx}
\usepackage{dcolumn}
\usepackage{bm}
\usepackage{subfigure}

\begin{document}


\title {
Low-lying energy spectrum of the cerium dimer
}

\author {A. V. Nikolaev}

\affiliation{Frumkin Institute of Physical Chemistry and
Electrochemistry of RAS, Leninskii pr. 31, 119991, Moscow, Russia}

\affiliation{Skobeltsyn Institute of Nuclear Physics, Moscow
State University, Vorob'evy Gory 1/2, 119991, Moscow, Russia}

\date{\today}

\begin{abstract}
The electronic structure of Ce$_2$ is studied in a valence bond model with two $4f$ electrons localized at two cerium sites.
It is shown that the low-lying energy spectrum of the simplest cerium chemical bond is determined
by peculiarities of the occupied $4f$ states.
The model allows for an analytical solution which is discussed along with the numerical analysis.
The energy spectrum is a result of the interplay between the $4f$ valence bond exchange, the $4f$ Coulomb repulsion
and the spin-orbit coupling.
The calculated ground state is the even $\Omega=\Lambda=\Sigma=0$ level,
the lowest excitations situated at $\sim30$ K
are the odd $\Omega=\Lambda=\Sigma=0$ state and the $^{3} 6_5$ doublet $(\Omega=\pm5,\, \Lambda=\pm6,\, \Sigma=\mp1)$.
The calculated magnetic susceptibility
displays different behavior at high and low temperatures.
In the absence of the spin-orbit coupling the ground state is the ${}^3 \Sigma_g^-$ triplet.
The results are compared with other many-electron calculations and experimental data.
\end{abstract}


\maketitle

\section{Introduction}
\label{sec:int}

Understanding the electronic nature of chemical bonds of $f$-elements is an issue
which has been attracting wide attention of investigators working in diverse areas of research.
The main question is the involvement of 4$f$ states in the formation of chemical bonds in molecules and in solids.
In general, one can speak of two extreme scenarios. That is, either $f-$electrons participate in chemical bonding which is
the case for 5$f$-states in the U$_2$ molecule \cite{Gag},
or they stay essentially localized at each site forming core-like states like in lanthanides.
This phenomenon known as the localization-delocalization paradigm in solid-state physics \cite{Alb}
is the subject of intense theoretical considerations and discussion.

In various compounds cerium is believed to display both types of 4$f$ behavior which makes it the key element in these studies.
Indeed, in solids Ce and its compounds
exhibit a broad range of unusual electronic, magnetic and structural properties called correlation effects \cite{Ful}.
While much effort has been invested in understanding of the complex electronic behavior in sophisticated cerium compounds,
very little attention has been given to the simplest cerium chemical bond \cite{Cao,Roo,She}, which is realized in Ce$_2$.
Meanwhile, owing to the progress of the spectroscopy of small metal clusters \cite{Lom}
precise experimental data on the cerium dimer \cite{She} have been reported recently,
followed by first reliable many-electron calculations \cite{Cao,Roo}.

Cao and Dolg \cite{Cao} have performed a series of calculations of Ce$_2$ using the complete active space self-consistent field (CASSCF) method,
multireference configuration interaction (MRCI) method and coupled cluster approach [CCSD(T)] and concluded that its ground state ``superconfiguration" is
$4f^1 4f^1$$(6s \sigma_g)^2 (5d\pi_u)^4$.
Six low-lying molecular terms ($^1 \Sigma_g^+$, $^1 \Sigma_u^-$, $^3 \Sigma_g^-$, $^3 \Sigma_u^+$, $^1 6_g$ and $^1 6_u$) are
``virtually degenerate" so that the exact refinement of the ground state is very difficult.
The authors then restricted themselves to the $^1 \Sigma_g^+$ and $^3 \Sigma_g^+$ ground states for which
in the scalar relativistic approximation they obtained the following spectroscopic constants:
$R_e=2.62 \pm 0.02$ {\AA}, $D_e=1.60 \pm 0.41$ eV and $\omega_c=201 \pm 13$ cm$^{-1}$.
The low-lying excitations were attributed to the core-like properties of $4f$ states. In the spin-orbit calculation
the authors obtained the $[\Omega=0]_g^+$ ground state followed by two low-lying levels $0_u^-$ ($\le$ 1 cm$^{-1}$) and $5_u$ ($\le$ 2 cm$^{-1}$).

The nature of the Ce$_2$ bonding and the role of the 4$f$-states were also considered in CASSCF studies by Roos {\it et al.} in \cite{Roo}.
As in \cite{Cao} it was found that the triple bond molecular configuration $(\sigma_g)^2 (\pi_u)^4$ couples in the $^1 \Sigma_g^+$ molecular term.
Two remaining nonbonding and
localized 4$f$ electrons  are in $\phi$-state ($m=\pm3$).
These two 4$f$ electrons then can form the configurations
$(4f \phi_g)^2$, $(4f \phi_g)(4f \phi_u)$, and $(4f \phi_u)^2$,
which result in molecular terms with angular momenta $L_z=0$ and 6.
The six calculated low-lying levels again are almost degenerate having approximately the same
spectroscopic parameters (without the spin-orbit interaction): $R_e=2.63-2.66$ {\AA}, $\omega_c=166-189$ cm$^{-1}$, $D_e=2.61-2.68$ eV
(The CASPT2 level of theory and a basis set of VQZP quality).
The spin-orbit coupling has only small effect on the computed properties, in particular it reduces $D_e$ to 2.51 eV in good agreement with
the experimental estimate $D_0=2.57$ eV [from the third law].
Calculations of Roos {\it et al.} give a longer $R_e$ and a softer $\omega_c$ in comparison with the values of Cao and Dolg \cite{Cao}.
Both theoretical estimates of $\omega_c$ \cite{Cao,Roo} are smaller than the experimental result \cite{She}:
$\omega_e=245.4 \pm 4.2$ cm$^{-1}$ [the force constant $k_e=2.28(1)$ mdyne/{\AA}].

Thus, the standard well-known many-electron procedures \cite{Cao,Roo} correctly describe the Ce$_2$ chemical bonding
but find six nearly degenerate low-lying molecular terms ($^1 \Sigma_g^+$, $^1 \Sigma_u^-$, $^3 \Sigma_g^-$, $^3 \Sigma_u^+$, $^1 6_g$ and $^1 6_u$)
instead of a well defined ground state. The difficulty lies
in the smallness of the splitting energies of these excitations requiring a special treatment.
In the following we formulate a simplified valence bond (VB) model accounting for the 4$f$ electrons in the cerium dimer,
find the important interactions and obtain the ``fine" energy spectrum of the electron excitations.
Our approach is based on the picture of 4$f$-4$f$ interactions as rightly found in \cite{Cao,Roo},
but requires a more deep insight in the couplings which experience the 4$f$ electrons in the molecule.

\section{model}
\label{sec:model}

We start with the description of the triple bond of Ce$_2$. The ground state electronic configuration of the cerium atom is $(4f)^1 (5d)^1 (6s)^2$
with the $^1 G_4$ ground term. This implies that the Ce atom is a rare case of the first Hund rule violation \cite{Mor}.
The nature of such behavior lies in ``quartic Fermi hole" which effectively separates the valence electrons and decreases their Coulomb repulsion \cite{Mor}.
Already at this stage of the investigation it is clear that the 4$f$ electron is not independent, through the Coulomb repulsion
it is strongly coupled with the valence 5$d$ and 6$s$ electrons on the same cerium site.

However, the $(4f)^1 (5d)^1 (6s)^2$ configuration with the closed $6s-$shell is not favorable for chemical bonding.
Therefore the electronic atomic state of Ce is first transformed to the $(4f)^1 (5d)^2 (6s)^1$ configuration with
the $^5 H_3$ term situated at 0.299 eV above the $^1 G_4$ ground term \cite{NIST}.
In the $(4f)^1 (5d)^2 (6s)^1$ configuration both 5$d$ and 6$s$ shells become open and form the triple bond with six valence electrons:
$(6s \sigma_g)^2 (5d\pi_u)^4$.
The $^1 \Sigma_g^+$ bonding state of 5$d$ and 6$s$ electrons (without the 4$f$ states) is well understood \cite{Cao,Roo},
in particular it occurs in La$_2$ where the molecular characteristics ($D_0=2.5$ eV, $k_e=2.28(1)$ mdyne/{\AA} \cite{She},
the $(5d)^1 (6s)^2$ initial and $(5d)^2 (6s)^1$ final [at 0.33 eV] configuration) are very close to those in Ce$_2$.
Although the triple bond is strong, the total energy gain is weakened by 0.6 eV because of the necessity to promote one 6$s$-electron
to the 5$d$-shell for both cerium atoms.

We consider the first cerium atom at (0,0,0) and the second at (0,0,$R_e$) (so that the $z$-axis is the molecular axis)
and find the valence density distribution of the four 5$d$ and two 6$s$ valence electrons.
The $^1 \Sigma_g^+$ $5d-6s$ ground state implies an axially symmetric electron density.
The conclusion can also be proved by considering the density distributions of two $s$ and four $d$-electrons
in the $(6s \sigma_g)^2 (5d\pi_u)^4$ configuration.

Now to the $5d$ and $6s$ valence states we add two $4f$-electrons.
The main assumption supported by \cite{Roo,Cao} is that to a first approximation
the 4$f$ electrons do not contribute to the chemical bonding and remain localized.
Then it follows that each 4$f$ electron should be either in the $Y_{l=3}^{3,c}(\Theta,\varphi)=C \sin^3\Theta \cos3\varphi$ or
in the $Y_{3}^{3,s}(\Theta,\varphi)=C \sin^3\Theta \sin3\varphi$ angular orbital state where $C=-\sqrt{70/\pi}/8$.
(Here we deal with the real spherical harmonics $Y_{l}^{m,t}$ ($t=c$ for cosine- and $t=s$ for sine-like azimuthal ($\varphi$-) dependence)
defined in Ref.\ \cite{BC}.)
That is because in both these states the Coulomb intrasite repulsion between the $4f$ electron and
the bonding $5d$ and $6s$ states at the cerium site is minimal (see Appendix A).
Indeed, in these $m=\pm3$ $f-$states the angular density dependence is $\sin^6 \Theta$ with the maximum at $\Theta=\pi/2$, while
in the binding $5d$ states maximum lies at $\Theta=\pi/4$ and in $6s$ states at $\Theta=0$ (considering their overlap).
Thus, the occupation of the $m=\pm3$ $f-$states ensures a certain space separation of the populated $4f$ states
from the binding $5d$ and $6s$ electrons, which effectively minimizes the intrasite $4f-6s$ and $4f-6s$ Coulomb repulsions.
The situation is discussed in detail in Appendix A, notice that the conclusion about the occupancy of the $4f \phi$ states is
also explicitly supported by the calculations of Roos et al. \cite{Roo} and implicitly by Cao and Dolg \cite{Cao}.

\begin{figure}
\resizebox{0.4\textwidth}{!} {
\includegraphics{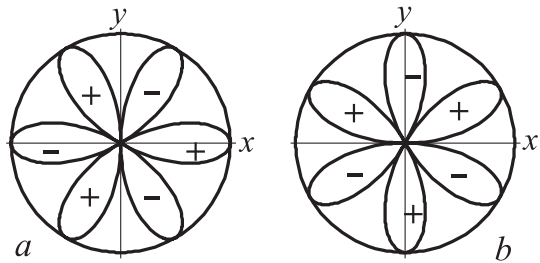}   }

\caption{ The azimuthal dependence of two $4f$ orbital functions: (a) cos-like $Y_{l=3}^{3,c}$, (b) sin-like $Y_{l=3}^{3,s}$.
The $z$-axis which is perpendicular to the plane of the picture (not shown) is parallel to the Ce$_2$ molecular axis.
} \label{fig1}
\end{figure}

Thus, at each cerium site the $4f$ electron space is limited to two $m=\pm3$ orbitals, see Fig.\ \ref{fig1}:
\begin{subequations}
\begin{eqnarray}
     & &\psi_{f,c}(\Theta,\varphi,r) = R_f(r) Y_{l=3}^{3,c}(\Theta,\varphi), \label{f1a} \\
     & &\psi_{f,s}(\Theta,\varphi,r) = R_f(r) Y_{l=3}^{3,s}(\Theta,\varphi) .
\label{f1b}
\end{eqnarray}
\end{subequations}
Here $R_f(r)$ is the radial part of the $4f$ states.
%
%

Since there is very little overlap between $4f$ functions belonging to different sites,
the valence bond (VB or Heitler-London) \cite{HL,VB} approach should give the best description to the $4f$ electronic problem.
Its basis set consists of the two electron $4f$ states:
\begin{eqnarray}
     \Psi_{o_a,o_b}(\vec{r}_1, s_{1z};\; \vec{r}_2, s_{2z}) =\psi_{f,o_a}(\vec{r}_1) u_{1s}\; \psi_{f,o_b}(\vec{r}_2) u_{2s} ,
\label{f2}
\end{eqnarray}
where the indices $o_a$ and $o_b$ refer to two orbital sites at sites $a$ and $b$, i.e.  $o_a,o_b=c$ or $s$ (orbital functions defined in Eq.\ (\ref{f1a},b) and
visualized in Fig.\ \ref{fig1}),
$s_z$ is the spin projection number ($\pm1/2$),
$u_s$ is the corresponding spin function ($\alpha$ or $\beta$).

The two electron wave function $\Psi_{o_a,o_b}(\vec{r}_1, s_{1z};\; \vec{r}_2, s_{2z})$=$\Psi_{O}(\vec{r}_1,\, \vec{r}_2)\, \Psi_{S}(s_{1z};\; s_{2z})$
is a product of the orbital part $\Psi_{O,i}$ ($i=1-4$), which can be labeled as
\begin{subequations}
\begin{eqnarray}
   |c,s\rangle, \quad |s,c\rangle, \quad |c,c\rangle, \quad |s,s\rangle ,
\label{f2a}
\end{eqnarray}
and the spin function $\Psi_{S,j}$ ($j=1-4$),
\begin{eqnarray}
   (\alpha,\alpha), \quad (\alpha,\beta), \quad (\beta,\alpha), \quad (\beta,\beta) .
\label{f2b}
\end{eqnarray}
\end{subequations}
(Here we omit cerium site indices $a$ and $b$, that is $|c,s\rangle=|c_a,s_b\rangle$ etc.)
Therefore, in total there are 16 two electron basis states $t$ ($t=$1-16), which are then antisymmetrized in respect to the electron permutation.

In atomic units the corresponding valence bond Hamiltonian \cite{HL,VB} is given by
\begin{subequations}
\begin{eqnarray}
   H^{VB} = -\frac{1}{2} \nabla^2_1 -\frac{1}{2} \nabla^2_2 +U(\vec{r}_1) + U(\vec{r}_2) + \frac{1}{r_{12}} + H^{so} , \nonumber \\
\label{h1a}
\end{eqnarray}
where $H^{so}$ is the spin-orbit coupling considered in detail below (Sec.\ \ref{subsec-so}), while
\begin{eqnarray}
   U(\vec{r}) = U_a(\vec{r}) + U_b(\vec{r}) .
\label{h1b}
\end{eqnarray}
\end{subequations}
Here $U_{a(b)}(\vec{r})<0$ is an effective $4f$ attraction to the nucleus $a$ (or $b$) screened by all core electrons.

It is well known that the evaluation of the matrix elements of $H^{VB}$ in the space of the functions (\ref{f2}) leads to the
following equation:
\begin{eqnarray}
   \langle t_1 | H^{VB} | t_2 \rangle = E_0\, \delta_{t_1,t_2} + J_{t_1,t_2} + K_{t_1,t_2} .
\label{h1c}
\end{eqnarray}
Here $E_0=E[(6s \sigma_g)^2 (5d\pi_u)^4]$ is the background molecular energy, $\delta$ is the Kronecker delta,
while $J$ and $K$ are the direct and exchange matrix elements
which we consider in Sec.\ \ref{subsecA} and Sec.\ \ref{subsecB}, respectively.

Taking into account the orbital degeneracy of 4$f$ states the matrix elements of $H$ (16$\times$16) where $H=J$, $K$ or $H^{so}$
are conveniently written in the spin block form,
\begin{eqnarray}
     H = \begin{pmatrix} H_{11} & H_{12} & H_{13} & H_{14} \\
                              H_{21} & H_{22} & H_{23} & H_{24} \\
                              H_{31} & H_{32} & H_{33} & H_{34} \\
                              H_{41} & H_{42} & H_{43} & H_{44}  \end{pmatrix} ,
\label{f3a}
\end{eqnarray}
where the indices $s_1,s_2$ refer to the spin part, Eq.\ (\ref{f2b}). As usual, $H_{21}=H_{12}^{\dagger}$ etc.
Each of the blocks in Eq.\ (\ref{f3a}) can be further represented as a 4$\times$4 orbital matrix of the following structure:
\begin{eqnarray}
     H_{s_1,\, s_2} = \begin{pmatrix} a_1  & c_1  & d_1 & e_1 \\
                                      c_1  & a_1  & d_2 & e_2 \\
                                      d_1^*& d_2^*& a_2 & c_2 \\
                                      e_1^*& e_2^*& c_2 & a_2  \end{pmatrix} .
\label{f3}
\end{eqnarray}
Here the matrix elements are evaluated between the two $f-$orbital sets [$i_1,i_2$, Eq.\ (\ref{f2a})] with the fixed spin indices [$s_1,s_2$, Eq.\ (\ref{f2b})].
In our basis with real spherical harmonics, $a_i$, $b_i$, $c_i$ are real,
while $d_i$ and $e_i$ are imaginary values describing the interaction with an external magnetic field
or the spin-orbit coupling.
The specific form of the matrix elements is considered below.

\subsection{$4f-4f$ direct Coulomb repulsion}

\label{subsecA}

Here we consider the matrix of the direct Coulomb repulsion:
\begin{eqnarray}
     J_{t_1,t_2} = \int \psi_{o_1,a}(\vec{r}_1) \psi_{o_3,a}(\vec{r}_1) \, V(\vec{r}_1,\vec{r}_2) \nonumber \\
                   \psi_{o_2,b}(\vec{r}_2) \psi_{o_4,b}(\vec{r}_2)\, d\vec{r}_1 d\vec{r}_2 , \label{h2}
\end{eqnarray}
where $V(\vec{r}_1,\vec{r}_2)=1/|\vec{r}_1 - \vec{r}_2|$ and the orbital indices
$(o_1,o_2)$ belong to the composite index $t_1$, while $(o_3,o_4)$ to $t_2$, Eq.\ (\ref{f2}).

From symmetry arguments it can be shown (Appendix B) that there are three types of the matrix elements,
\begin{eqnarray}
     A = \langle c, s | V | c, s \rangle = (c_a,c_a | V | s_b,s_b ) = (s_a,s_a | V | c_b,c_b ), \nonumber \\
     B = \langle c, c | V | c, c \rangle = (c_a,c_a | V | c_b,c_b ) = (s_a,s_a | V | s_b,s_b ), \nonumber  \\
     C = \langle c, s | V | s, c \rangle = (c_a,s_a | V | c_b,s_b ) = (c_a,s_a | V | s_b,c_b ), \nonumber  \\
\label{f4}
\end{eqnarray}
while
\begin{eqnarray}
     \langle c, c | V | c, s \rangle = (c_a,c_a | V | c_b,s_b ) = (s_a,s_a | V | c_b,s_b ) = 0 . \nonumber
\end{eqnarray}
Here we use both standard and so called chemist's notations \cite{Sza}. However, it turns out that not all parameters in (\ref{f4}) are
independent. Analyzing the azimuthal dependence of the integrands and ``the azimuthal selection rule" (Appendix B), one finds
\begin{eqnarray}
     B - A = 2 C . \label{f5a}
\label{f5b}
\end{eqnarray}

In the $H=J$ case in Eq.\ (\ref{f3a}) only the diagonal blocks $J_{ss}$ are relevant ($J_{s_1,s_2} \ne 0$ for $s_1=s_2$) because
the matrix elements (\ref{h2}) are independent of the spin indices.
Furthermore, the $J_{ss}$ blocks are identical, i.e. $J^{orb}=J_{ss}$, Eq.\ (\ref{f3}).
The $J^{orb}$ matrix is given by Eq.\ (\ref{f3}) where $a_1=A$, $a_2=B$, $c_1=c_2=C$ and $d_1=d_2=e_1=e_2=0$.

\subsection{$4f-4f$ exchange interaction}

\label{subsecB}

For the matrix element of the $4f-4f$ exchange interaction we obtain
\begin{eqnarray}
     K_{t_1,\, t_2} = -(S_{o_1,\, o_4}^{a,b}\, U^{a,b}_{o_2,\, o_3} + U_{o_1,\, o_4}^{a,b}\, S^{a,b}_{o_2,\, o_3}  \nonumber \\
     + V^{exc}_{o_1,o_4;o_2,o_3})\,  \delta_{s_{1z},\, s_{4z}}\, \delta_{s_{2z},\, s_{3z}} ,
\label{f8a}
\end{eqnarray}
where the composite indices $t_1$ and $t_2$ refer to $(o_{1,a},s_{1z};\, o_{2,b},s_{2z})$ and $(o_{3,a},s_{3z};\, o_{4,b},s_{4z})$ as
described earlier, Eq.\ (\ref{f2}), and
\begin{subequations}
\begin{eqnarray}
     S_{o_1,o_2}^{a,b} = \int d\vec{r}\, \psi_{o_1,a}(\vec{r}) \, \psi_{o_2,b}(\vec{r}), \label{f8aa} \\
     U_{o_1,o_2}^{a,b} = \int d\vec{r}\, \psi_{o_1,a}(\vec{r}) \, U(\vec{r}) \, \psi_{o_2,b}(\vec{r}) . \label{f8b}
\end{eqnarray}
Finally, the Coulomb exchange integral in (\ref{f8a}) is given by [compare with $J_{t_1,t_2}$ in Eq.\ (\ref{h2})]
\begin{eqnarray}
     V^{exc}_{o_1,o_4;o_2,o_3} = \int \psi_{o_1,a}(\vec{r}_1) \psi_{o_4,b}(\vec{r}_1)\, V(\vec{r}_1,\vec{r}_2)            \nonumber   \\
                                        \psi_{o_2,b}(\vec{r}_2) \psi_{o_3,a}(\vec{r}_2) d\vec{r}_1 d\vec{r}_2 .
\label{f8c}
\end{eqnarray}
\end{subequations}
If two first terms on the right hand side of (\ref{f8a}) differ from zero then they are leading, while the third (Coulomb) term is just a minor correction.

In our case there are two $4f$ functions at each site, Eq.\ (\ref{f1a})-(\ref{f1b}), but
from symmetry it follows that in all overlap integrals $S_{o_1,o_2}^{a,b}$ there is only one nonzero value,
\begin{eqnarray}
     S = \langle c_a | c_b \rangle = \langle s_a | s_b \rangle > 0,
\label{f8}
\end{eqnarray}
while $\langle c_a | s_b \rangle = 0$.
Analogously, $U_{exc}=\langle c_a | U_a | c_b \rangle=\langle s_a | U_a | s_b \rangle$ and $\langle c_a | U_a | s_b \rangle=0$.
It is convenient to introduce the notation
\begin{eqnarray}
     \delta E = -2S\, U_{exc} > 0
\label{h3}
\end{eqnarray}
for the leading term.

For the exchange Coulomb interaction $V^{exc}$, Eq.\ (\ref{f8c}),
one finds three types of integrals with nonzero values (compare with Eqs.\ (\ref{f4})):
\begin{subequations}
\begin{eqnarray}
     A' = \langle c, s | V | c, s \rangle_{exc} = (c_a,c_b | V | s_a,s_b ), \label{f11a} \\
     B' = \langle c, c | V | c, c \rangle_{exc} = (c_a,c_b | V | c_a,c_b ), \label{f11b} \\
     C' = \langle c, s | V | s, c \rangle_{exc} = (c_a,s_b | V | c_a,s_b ). \label{f11c}
\end{eqnarray}
\end{subequations}
Furthermore, there is the same relation as for the direct matrix elements $A$, $B$ and $C$, Eq.\ (\ref{f5a}), see Appendix B:
\begin{eqnarray}
     B' - A' = 2 C', \label{f12}
\end{eqnarray}

Then, in the full 16$\times$16 matrix $H=K$, Eq.\ (\ref{f3a}), the exchange coupling is effective in the blocks
$K_{11}^{exc}$ (between states $\alpha$, $\alpha$), $K_{23}^{exc}$ and $K_{32}^{exc}$
(between the states with $\alpha$, $\beta$ and $\beta$, $\alpha$) and $K_{44}^{exc}$ (between states $\beta$, $\beta$).
Each of the blocks has the following (orbital) form:
\begin{eqnarray}
     K_{s_1,\, s_2}^{exc} = \begin{pmatrix} -C' & \delta E - A' & 0 & 0 \\
                        \delta E - A' & -C' & 0 & 0 \\
                                    0 & 0 & \delta E - B' & -C' \\
                                    0 & 0 & -C' & \delta E - B'  \end{pmatrix} .
   \label{f12'}
\end{eqnarray}
(Here as in Ref. \cite{HL} we have taken into account the effect of the half canceling of the $\delta E$-terms 
in the final expression
due to the nonorthogonality of the $f$-functions.)

\subsection{Spin-orbit coupling}

\label{subsec-so}

Now in the framework of our model we consider the $4f$ spin-orbit interaction:
\begin{eqnarray}
     H^{so} =
     \zeta \, \vec{L}_1 \vec{s}_1 + \zeta \, \vec{L}_2 \vec{s}_2 ,
\label{f20}
\end{eqnarray}
where $\zeta = \zeta_f \approx 862$~K. This value corresponds to the $\triangle_{so}=(7/2)\zeta=$0.26 eV splitting between $j=5/2$ and $7/2$
one electron atomic $4f$ states.
(The energies of the states were obtained by solving the spherically symmetric Dirac equation for atomic cerium states in the framework
of the self-consistent field DFT-LDA atomic model.)
In our model the $4f$ quantum space is reduced to two orbital states with $m_z=\pm3$ at each site, and $H^{so}$ transforms to
\begin{eqnarray}
   H^{so}_z=\zeta \, L_{1,z} s_{1,z} + \zeta \, L_{2,z} s_{2,z} .
\label{f21}
\end{eqnarray}
Therefore, the nonzero spin blocks of the full matrix are $H_{s,s}^{so}$ ($s=1-4$), Eq.\ (\ref{f3a}), with the following general structure of each block
(see Appendix C):
\begin{eqnarray}
     H_{s, s}^{so} = \frac{3}{2} \zeta \begin{pmatrix} 0 & 0 & n_1 i & n_2 i \\
                                      0 & 0 & m_1 i & m_2 i \\
                                      -n_1 i & -m_1 i & 0 & 0 \\
                                      -n_2 i & -m_2 i & 0 & 0  \end{pmatrix} .
\label{f14}
\end{eqnarray}
In the $H_{11}^{so}$ block $n_1=m_1=1$, $n_2=m_2=-1$, in the $H_{22}^{so}$ block $n_1=n_2=-1$, $m_1=m_2=1$,
in the $H_{33}^{so}$ block $n_1=n_2=1$, $m_1=m_2=-1$ and in $H_{44}^{so}$ block $n_1=m_1=-1$, $n_2=m_2=1$.

The spin-orbit coupling, Eq.\ (\ref{f21}), splits the four single electron states at each cerium site in two doublets:
the lowest level with $j_z=\pm 5/2$ and the energy $E(\pm5/2)=-3\zeta/2$ has the eigenfunctions $\psi_{5/2,+} \sim \exp(3mi \varphi) \beta$,
$\psi_{5/2,-} \sim \exp(-3mi \varphi) \alpha$. The other level with $j_z=\pm 7/2$ and the energy $E(\pm7/2)=+3\zeta/2$ has
the functions $\psi_{7/2,+} \sim \exp(3mi \varphi) \alpha$, $\psi_{7/2,-} \sim \exp(-3mi \varphi) \beta$.
Correspondingly, two 4$f$ electrons have the following energy spectrum:
$E_{so,1}=-3\zeta$ (4 degenerate states), $E_{so,2}=0$ (8 states), $E_{so,3}=3\zeta$ (4 states).
The inclusion of other interactions modifies the picture.


\section{Results and discussion}
\label{sec:res}

All interaction parameters of the model introduced in Sec. \ref{sec:model} were evaluated numerically with our valence bond program \cite{NM5}, Table \ref{table1}.
There, we have used the overlapping cerium relativistic atomic $4f$ functions (the large $j=5/2$ component for $R_f(r)$)
calculated within the density functional theory (DFT)
in the local density approximation (LDA) \cite{DFT}.
%
\begin{table}
\caption{
Calculated $4f$ direct ($A,B,C$) and exchange ($A',B',C'$) Coulomb integrals,
the overlap integral $S$, the VB attraction $U_{exc}$
and the $4f$ exchange parameter $\delta E$ for various Ce-Ce distances $R$ (in {\AA}).
All energy parameters are in Kelvin~(K). \\
$^a$ Ce$_2$ equilibrium distance for $^1 \Sigma_g^+$, $^3 \Sigma_u^+$ in Ref.\ \cite{Cao}, \\
$^b$ Ce$_2$ equilibrium distance for $^1 \Sigma_g^+$, $^1 6_g$, $^3 \Sigma_u^+$, $^3 6_u$ in Ref.\ \cite{Roo}, \\
$^c$ Ce$_2$ equilibrium distance for $^1 \Sigma_u^-$, $^3 \Sigma_g^-$ in Ref.\ \cite{Roo}. \\
$R=$3.43 {\AA} and 3.65 {\AA} correspond to the nearest neighbor distance in $\alpha$-Ce and $\gamma$-Ce \cite{Kosk},
respectively.
\label{table1}     }

\begin{ruledtabular}
 \begin{tabular}{ c  c c c  c c c }
            & $R=$2.62$^a$  & 2.63$^b$ & 2.66$^c$ & 2.70 & 3.43 ($\alpha$) & 3.65 ($\gamma$) \\
\hline
     $B$    &  2324.45 &  2286.28 &  2174.68 &  2032.49 &   643.73 &  471.58 \\
     $A$    &  2323.98 &  2285.82 &  2174.26 &  2032.12 &   643.69 &  471.56 \\
     $C$    &     0.24 &     0.23 &     0.21 &     0.19 &     0.02 &    0.01 \\

     $B'$   &     1.82 &     1.75 &     1.56 &     1.34 &     0.09 &    0.04 \\
     $A'$   &     1.67 &     1.61 &     1.43 &     1.23 &     0.09 &    0.04 \\
     $C'$   &     0.07 &     0.07 &     0.06 &     0.05 &        0 &       0 \\

$S \times 10^3$& 4.224 &    4.148 &    3.928 &    3.654 &    1.029 &   0.714 \\
     $U_{exc}$    & -1727.81 & -1688.96 & -1578.02 & -1442.24 &  -309.24 &  -200.34 \\
 $\delta E$ &    -14.6 &   -14.01 &    -12.4 &   -10.54 &    -0.64 &    -0.28
 \end{tabular}
\end{ruledtabular}
\end{table}

Below we discuss the results of the valence bond model with ($H^{so} \neq 0$) and without ($H^{so} = 0$) the spin-orbit coupling.

\subsection{VB without the spin-orbit coupling}

From Table \ref{table1} we see that the largest parameters are $A$ and $B$
which correspond to the direct Coulomb repulsion, Sec.\ \ref{subsecA}.
Therefore, as a first approximation we consider only the matrix of the direct Coulomb repulsion $J$.
In that case from the structure of the spin matrix it follows that the triplet and singlet states coincide.
Having diagonalized the corresponding 4$\times$4 orbital matrix analytically, we find its eigenvalues,
\begin{subequations}
\begin{eqnarray}
   & & E_1 - A = -C, \label{f6a} \\
   & & E_{2,3} - A = C \label{f6b} \\
   & & E_4 - A = 3C .
\label{f6c}
\end{eqnarray}
\end{subequations}
(The eigenvectors are quoted in Eqs.\ (\ref{c7a})-(\ref{c7c}), Appendix C.)
We conclude that the energy splittings are determined by the parameter $C$, Eq.\ (\ref{f4}), which is very small, while
$A$ should be incorporated into the background molecular energy $E_0=E[(6s \sigma_g)^2 (5d\pi_u)^4]$.

Notice that the eigenvalues $E_2$ and $E_3$ are degenerate which implies a nonzero orbital momentum.
In Appendix C we show that indeed the two components are attributed to $M_z=\pm 6$. Taking into account
other properties of the eigenvectors one can show that the molecular terms are identified as
$\Sigma_u^-$, $6_g$ and $\Sigma_g^+$, respectively. However, the presented picture is not full
because the integral $C$ is very small, Table \ref{table1}.
If $C$ were of the order of $A$, the direct Coulomb splitting would have been dominant,
but since $C \sim 0.2$~K we should consider other relevant interactions.

Now in addition to the direct Colomb repulsion $J$ we consider the $4f-4f$ exchange interaction $K$, Sec.\ \ref{subsecB}.
A small overlap $S$ between two relevant $4f$ functions splits the singlet and triplet states.
If there were only one orbital for each site, the spin singlet state would have been lower in energy than the spin triplet state.
Indeed, in the single orbital valence bond problem \cite{HL,VB} we have
\begin{eqnarray}
     & & E_S = E_0 - \delta E , \quad E_T = E_0 + \delta E , \label{f10a}
\label{f10b}
\end{eqnarray}
where $\delta E$ is given by Eq.\ (\ref{h3}).
In our case of two 4$f$ orbitals at each site the reverse is true.
This can be clearly seen if for the time being we ignore the exchange repulsion ($A'=B'=C'=0$),
and treat the electron site exchange as a perturbation.
As an example we consider the state (\ref{f6a}) with the eigenstate (\ref{c7a}).
Its orbital part can be written as $[-c_a(\vec{r}_{1a}) s_b(\vec{r}_{2b}) + s_a(\vec{r}_{1a}) c_b(\vec{r}_{2b})]/\sqrt{2}$,
where $\vec{r}_{1a}=|\vec{r}_1-\vec{R}_a|$, $\vec{r}_{2b}=|\vec{r}_2-\vec{R}_b|$.
In the spin singlet state we then symmetrize the orbital
part, obtain $[-c_a(\vec{r}_{1a}) s_b(\vec{r}_{2b}) - c_a(\vec{r}_{2a}) s_b(\vec{r}_{1b}) + s_a(\vec{r}_{1a}) c_b(\vec{r}_{2b})
+ s_a(\vec{r}_{2a}) c_b(\vec{r}_{1b})]/2$ and find $E(^1 \Sigma_u^-)=E_1+\delta E$.
In the triplet state we antisymmetrize the orbital part, obtain
$[-c_a(\vec{r}_{1a}) s_b(\vec{r}_{2b}) + c_a(\vec{r}_{2a}) s_b(\vec{r}_{1b}) + s_a(\vec{r}_{1a}) c_b(\vec{r}_{2b})
- s_a(\vec{r}_{2a}) c_b(\vec{r}_{1b})]/2$
and find $E(^3 \Sigma_g^-)=E_1-\delta E$.
Thus, the Ce$_2$ ground state is the ${}^3 \Sigma_g^-$ triplet state in contrast to the standard (diatomic molecule) case.
The origin of this anomalous behavior is most likely connected with the fact that the orbital part of the triplet state is even.
The results for all states are summarized in Table \ref{table2}, which holds even if $\delta E$ is not a perturbation.
The energy levels are also schematically shown in Fig.\ \ref{fig2}.
%
\begin{table}
\caption{
The VB energy spectrum of Ce$_2$ without the $4f$ spin-orbit coupling ($\zeta=0$)
and the exchange Coulomb repulsion ($A'=B'=C'=0$).
\label{table2}     }

\begin{ruledtabular}
 \begin{tabular}{ c c c c }
 singlets  & $E-E_0$  & triplets & $E-E_0$ \\
\hline
  $^1 \Sigma_u^-$ & $-C+\delta E$  & $^3 \Sigma_g^-$ &  $-C-\delta E$  \\
  $^1 6_g$        & $ C-\delta E$  & $^3 6_u$        &  $ C+\delta E$  \\
  $^1 \Sigma_g^+$ & $3C-\delta E$  & $^3 \Sigma_u^+$ &  $3C+\delta E$
 \end{tabular}
\end{ruledtabular}
\end{table}

Thus, the situation in Ce$_2$ could be an interesting example of the triplet ground state. In practice however the relatively strong
(in comparison with $\delta E$) spin-orbit coupling changes the electron spectrum.

\subsection{VB with the spin-orbit coupling}

In this section we discuss the most general case, when the full valence bond Hamiltonian $H^{VB}$ is given by Eq.\ (\ref{h1a}).

Surprisingly, an analytical solution can be figured out even for this case
except for the small Coulomb repulsion which accompanies the 4$f$ exchange transitions
[i.e. $A'=0$, $B'=0$, $C'=0$, Eq.\ (\ref{f11a})-(\ref{f11c})].
The analytical expressions for the energy levels are quoted in Table \ref{table3}.
%
\begin{table}
\caption{
The analytical expressions for the energy levels of Ce$_2$ in the VB model with the spin-orbit coupling  
($A'=0$, $B'=0$, $C'=0$), $P$ is the state parity.
\label{table3}     }

\begin{ruledtabular}
 \begin{tabular}{ c c c c c c c }
    & $E$  & deg.  &  $\Omega\, (J_z)$ & $\Lambda\, (L_z)$ & $\Sigma\, (S_z)$ & $P$  \\
1 &  $-\sqrt{9 \zeta^2+4C^2}+C-\delta E$ & 1 & 0 & 0 & 0 & $g$  \\
2 &  $-\sqrt{9 \zeta^2+4C^2}+C+\delta E$ & 1 & 0 & 0 & 0 & $u$  \\
3 &  $ -3\zeta + C + \delta E$ & 2 & $\mp5$ & $\mp6$ & $\pm1$ & $u$ \\
4 &  $ -C - \delta E$          & 2 & $\mp1$ & 0 & $\mp1$ & $g$ \\
5 &  $  C - \delta E$          & 2 & $\mp6$ & $\mp6$ & 0 & $g$ \\
6 &  $  C + \delta E$          & 2 & $\mp6$ & $\mp6$ & 0 & $u$ \\
7 &  $ 3C + \delta E$          & 2 & $\mp1$ & 0 & $\mp1$ & $u$ \\
8 &  $\sqrt{9 \zeta^2+4C^2}+C-\delta E$ & 1 & 0 & 0 & 0 & $g$ \\
9 &  $ 3\zeta + C + \delta E$ & 2 & $\mp7$ & $\mp6$ & $\mp1$ & $u$ \\
10&  $\sqrt{9 \zeta^2+4C^2}+C+\delta E$ & 1 & 0 & 0 & 0 & $u$

 \end{tabular}
\end{ruledtabular}
\end{table}
Notice that the three lowest states are those found in Ref.\ \cite{Cao} [$(\Omega=0)_g$, $0_u$ and $5_u$] and Ref.\ \cite{Roo} (Table 6).

For completeness we can include the 4$f$ electron exchange repulsion ($A'$, $B'$, $C'$) and
compute the energy levels numerically.
The calculated energy values for characteristic internuclear distances $R$ and the parameters of Table \ref{table1}
are quoted in Table \ref{table4}.
Since $A'$, $B'$ and $C'$ are very small, they introduce only
a minor correction to the analytical expressions.

Usually, for labeling of molecular states one uses the standard many-electron spectroscopic notation ${}^{2S+1} \Lambda_{\Omega}$ \cite{Lan}.
However, in the $\Sigma=S_z=0$ states the inclusion of the spin-orbit coupling mixes up singlet ($S=0$) with triplet ($S=1$) states
and the partial weights of singlet and triplet contributions in the final $S_z=0$ states are nearly equal. (The effect was found in \cite{Cao,Roo}.)
In our model for the lowest three states we obtain:
\begin{subequations}
\begin{eqnarray}
   & &\Psi_{1\,(gs)} =  a \Psi({}^3 \Sigma_g^-;\, S_z=0) - b \Psi({}^1 \Sigma_g^+;\, S_z=0), \quad \quad \label{mix1} \\
   & &\Psi_2 = -a \Psi({}^1 \Sigma_u^-;\, S_z=0) + b \Psi({}^3 \Sigma_u^+;\, S_z=0), \quad \quad \label{mix2} \\
   & &\Psi_{3,1} = -a \Psi({}^3 \Sigma_u^-;\, S_z=1) + b \Psi({}^3 \Sigma_u^+;\, S_z=1), \quad \quad \\
   & &\Psi_{3,2} =  b \Psi({}^3 \Sigma_u^-;\, S_z=-1) + a \Psi({}^3 \Sigma_u^+;\, S_z=-1). \quad \quad \quad
\end{eqnarray}
\end{subequations}
Here $a=0.7072$, $b=0.7070$, $\Psi({}^3 \Sigma_g^-;\, S_z=0)$ is the $S_z=0$ component of the ${}^3 \Sigma_g^-$ triplet etc.
Notice that the ${}^3 \Sigma_u^-$ and ${}^3 \Sigma_u^+$ states in $\Psi_3$ form the ${}^3 6_u$ level ($\Omega=5$),
see Fig.\ref{fig2} below.
In the following we will use the short notation $(1,3)$ for the spin multiplicity ($2S+1$) of the strongly mixed states.
For example, the ground state is labeled as ${}^{3,1} \Sigma_{0,g}$ (or ${}^{1,3} \Sigma_{0,g}$) etc.

The exchange Coulomb interaction ($A',B',C' \ne 0$) splits the doublets 5 and 6 ($\Omega=\Lambda=6$, $\Sigma=0$, $g,u$),
Table \ref{table3}:
\begin{eqnarray}
     {}^{1,3} 6_{6\,(g,u)} \rightarrow {}^{1,3} \Sigma_{0\,(g,u)}^+ + {}^{1,3} \Sigma_{0\, (g,u)}^- .
\label{d1}
\end{eqnarray}
In the splitted lines (5a, 5b, 6a, 6b in Table \ref{table4}) $\Omega = 0$, $\Lambda = 0$ and $\Sigma=0$.
Although the ordering of three lowest levels is the same as in Ref.\ \cite{Cao}, the energy splittings are
different. At $R=2.62$ {\AA} we have $E_2-E_1=E_3-E_1=32.7$ K, while in the spin-orbit calculation of
Cao and Dolg \cite{Cao} $E_2-E_1 \le 1$ cm$^{-1}$ and $E_3-E_1 \le 2$ cm$^{-1}$.

%
\begin{table}
\caption{
Calculated energy levels (in K) of Ce$_2$ at various characteristic internuclear distances
$R$ (in {\AA}) with the spin-orbit coupling, see also notes for Table \ref{table1}.
\label{table4}     }

\begin{ruledtabular}
 \begin{tabular}{ c  c c c  c c c }
            & $R=$2.62  & 2.63 & 2.66 & 2.70 & 3.43 ($\alpha$) & 3.65 ($\gamma$) \\
\hline
 1   & -2602.11 & -2601.46 & -2599.69 & -2597.64 & -2586.71 & -2586.31 \\
 2   & -2569.42 & -2570.08 & -2571.90 & -2573.99 & -2585.25 & -2585.67 \\
 3   & -2569.42 & -2570.08 & -2571.90 & -2573.99 & -2585.25 & -2585.67 \\
 4   & -16.59   &  -15.92  &   -14.11 &   -12.02 &    -0.75 &    -0.33 \\
 5$a$& -16.26   &  -15.60  &   -13.82 &   -11.75 &    -0.71 &    -0.31 \\
 5$b$& -15.96   &  -15.32  &   -13.56 &   -11.53 &    -0.71 &    -0.31 \\
 6$a$&  16.44   &   15.78  &    13.98 &    11.91 &     0.75 &     0.33 \\
 6$b$&  16.74   &   16.06  &    14.24 &    12.13 &     0.75 &     0.33 \\
 7   &  17.07   &   16.38  &    14.53 &    12.40 &     0.79 &     0.35 \\
 8   & 2569.90  &  2570.54 &  2572.32 &  2574.37 &  2585.29 &  2585.69 \\
 9   & 2602.59  &  2601.92 &  2600.11 &  2598.02 &  2586.75 &  2586.33 \\
 10  & 2602.59  &  2601.92 &  2600.11 &  2598.02 &  2586.75 &  2586.33
 \end{tabular}
\end{ruledtabular}
\end{table}

It is worth noting that the levels 2, 3 and 9, 10 are virtually degenerate.
According to Table \ref{table3} the energy difference between them is
$2C^2/3\zeta \ll (\zeta,\, \delta E)$.

If the spin-orbit coupling is dominant (which is the case for Ce$_2$), the
twelve energy levels are clearly divided in three groups:
\begin{subequations}
\begin{eqnarray}
      (1)\; & {}^{3,1} \Sigma_{0\,(g)},\, {}^{1,3} \Sigma_{0\, (u)}, \, {}^3 6_{5\, (u)} .       &  (\sim E_{so,1}) \quad  \\
      (2)\; & {}^3 \Sigma_{1\, (g,-)},\, {}^{1,3} \Sigma_{0\, (g,+)},\, {}^{1,3} \Sigma_{0\, (g,-)},      &  \nonumber \\
            &     {}^{1,3} \Sigma_{0\, (u,-)},\, {}^{1,3} \Sigma_{0\, (u,+)},\, {}^3 \Sigma_{1\, (u,+)} . &  (\sim E_{so,2}) \quad  \\
     (3)\;  & {}^{1,3} \Sigma_{0\, (g)},\, {}^3 6_{7\, (u)},\; {}^{1,3} \Sigma_{0\, (u)} .        &  (\sim E_{so,3}) \quad
\label{d3}
\end{eqnarray}
\end{subequations}
Here the levels are given in the increasing energy order as in Tables \ref{table3} and \ref{table4}
with some additional characteristics (the parity and the mirror reflection, when possible) in parentheses.
If the other interactions are omitted (i.e. $\delta E=0$, $C=0$, and $A'=B'=C'=0$) then the energy
spectrum is reduced to only three levels ($E_{so,1}=-3\zeta$, $E_{so,2}=0$, $E_{so,3}=+3\zeta$)
discussed in Sec.\ \ref{subsec-so}.

The relation between three important characteristic parameters of Ce$_2$ is the following
$\zeta \gg \delta E \gg C$.
However, the analytical and numerical solution can be used to study the model for
any parameter set. It is clear that such a scenario is not
relevant for Ce$_2$, but it can possibly occur in other electronic systems
with localized $f$ electrons. In particular, it is instructive to consider
the case of a small spin-orbit coupling: $(C,\, \delta E) \gg \zeta$.
For that relation (provided that $A'=B'=C'=0$)
one can still use the analytical solution, Table \ref{table3},
but of course the order and level grouping will be different.
The splittings and relations between energy levels in the absence of the spin-orbit coupling ($\zeta=0$)
are schematically shown in Fig.\ \ref{fig2}.
%
\begin{figure}
\resizebox{0.4\textwidth}{!} {
\includegraphics{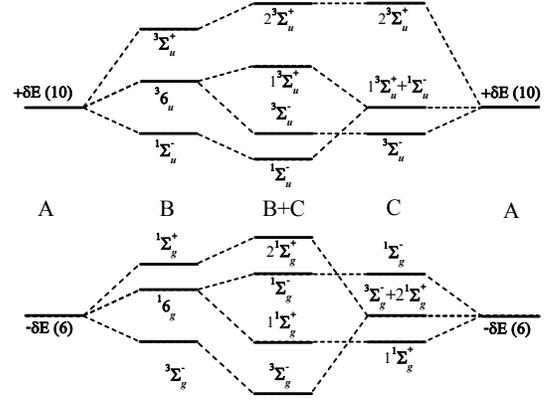}   }

\caption{ The schematic energy splittings in the VB model 
in the absence of the spin-orbit coupling: (A) only $\delta E \ne 0$;
(B) $\delta E,\, C \ne 0$, $A',B',C'=0$ ; (C) $\delta E,\, A',B',C' \ne 0$, $C=0$; (B+C) $\delta E,\, A,B,C,\, A',B',C' \ne 0$.
In the case (C) the levels ${^3}\Sigma_g^-$, ${^1}\Sigma_g^+$ and ${^3}\Sigma_u^+$, ${^1}\Sigma_u^-$
are ``accidentally'' degenerate.
} \label{fig2}
\end{figure}
%

In Fig.\ \ref{fig2} the most realistic subcase (all relevant interactions included) is (B+C).
If on top of it a small spin-orbit coupling is introduced, then
the spin triplet states there are split according to the following scheme:
\begin{subequations}
\begin{eqnarray}
     & & {}^3 \Sigma_g^- \rightarrow {}^3 \Sigma_{0\, (g,-)} + {}^3 \Sigma_{1\, (g,-)} , \\
     & & {}^3 \Sigma_u^- \rightarrow {}^3 \Sigma_{0\, (u,-)} + {}^3 6_{5\, (u)} , \\
     & & 1\, {}^3 \Sigma_u^+ \rightarrow {}^3 \Sigma_{0\, (u,+)} + {}^3 6_{7\, (u)} , \\
     & & 2\, {}^3 \Sigma_u^+ \rightarrow {}^3 \Sigma_{0\, (u,+)} + {}^3 \Sigma_{1\, (u,+)} .
\label{d2}
\end{eqnarray}
\end{subequations}

\section{Conclusions}
\label{sec:con}

For the cerium dimer we proposed a model of two interacting $4f$ electrons which
can explain and refine the low-lying energy spectrum of Ce$_2$ observed in
many electron computational studies \cite{Cao,Roo}. The model is a modified valence bond approach where
two relevant $4f$ orbital states at each cerium site are explicitly taken into account,
Eqs.\ (\ref{f1a}), (\ref{f1b}), and Fig.\ \ref{fig1}.
The other valence $6s$ and $5d$ states are grouped in a conventional triple
chemical bond with the $(6s \sigma_g)^2 (5d\pi_u)^4$ closed molecular shell.

In the framework of the $4f$ model we can clearly trace the origin
of the energy levels and establish intrinsic relations between them.
For the most important interactions the problem of finding the $4f$ energy spectrum
is solved analytically.
Apart from the spin-orbit coupling ($\zeta =$862 K) the largest interaction is the direct Coulomb
repulsion of the two localized $4f$ electrons (described by parameters $A,B$)
but the effective Coulomb parameter responsible for the energy
splitting is very small ($C \sim $0.2 K), Table \ref{table1}.
The next important interaction is the valence bond
exchange of the two localized $4f$ electrons ($\delta E \sim$14 K) which is also accompanied
by small exchange Coulomb repulsion ($A',B' \sim$1.5 K, $C' \sim$0.07 K).

\begin{figure}
\resizebox{0.4\textwidth}{!} {
\includegraphics{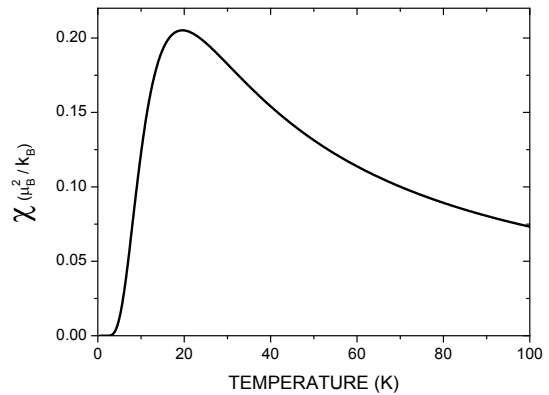}   }

\caption{ The calculated magnetic susceptibility of the cerium dimer.
The magnetic field is parallel to the molecular axis. The Curie law tail
at high temperatures can be fit with the effective Bohr magneton number
2.42 $\mu_B$ (about 300 K) per cerium site.
} \label{fig3}
\end{figure}

As a result, the even ground state level with $\Omega=\Lambda=\Sigma=0$
(composed of ${}^3 \Sigma_{0\,(g)}$ and ${}^1 \Sigma_{0\, (g)}$ states, Eq.\ (\ref{mix1}))
is separated from first excited levels with $\Omega=\Lambda=\Sigma=0$ (composed of
${}^3 \Sigma_{0\,(u)}$ and ${}^1 \Sigma_{0\, (u)}$ states, Eq.\ (\ref{mix2})) and  ${}^3 6_{5\, (u)}$ by an energy gap
of $\sim$30 K.
The ordering of these levels is in agreement with calculations of Cao and Dolg \cite{Cao},
although the energy splittings are different.
The $J_z=\Omega=\pm5$ excited states support magnetic moments, giving rise to the magnetic
susceptibility dependence $\chi$ shown in Fig.\ \ref{fig3}. The magnetic susceptibility
clearly displays two different regimes. At large temperatures ($T > 30$ K) it follows the Curie law
indicative of ``free" magnetic moments. At small temperatures ($T < 20$ K)
$\chi$ deviates from the Curie law and rapidly decreases as $T \rightarrow 0$.
Such low $T$ behavior and disappearance of the ``free" $4f$ magnetic moments of cerium
is caused by the population of the nonmagnetic ground state ($\Omega=\Lambda=\Sigma=0$)
at the expense of the excited magnetic ones.
The conclusion can be checked by experimental measurements.

We have also studied the energy spectrum in the absence of the spin-orbit coupling,
Fig.\ \ref{fig2}. Here, the unusual feature is the ${}^3 \Sigma_g^-$ triplet ground state.

The peculiarities of the electronic structure of the simplest cerium bond possibly can
clarify the picture of phase transitions in solid cerium \cite{Tsv}.

%
%

I am grateful to K.H. Michel, B.N. Plakhutin, A.V. Tsvyashchenko, E.V. Tkalya, A.V. Avdeenkov and B. Verberck
for valuable discussions.

\appendix

\section{Appendix A}

According to the recent many electron calculations (CASSCF) of Ce$_2$ \cite{Roo} each cerium site has a single $4f$ electron which is virtually nonbonding
and preserves its localized nature. Furthermore, without the spin-orbit coupling it occupies one of the two orbitally degenerate $4f$ states
with the highest azimuthal angular momentum ($m=\pm3$). Although it is not mentioned explicitly, the same orbital degeneracy follows from the low-lying molecular term
spectrum computed by Cao and Dolg \cite{Cao}. Below we show that the lowering of the $m=\pm3$ states is caused by the Coulomb repulsion between the $4f$ states and
the bonding electron states.

We consider a simplified model with two $5d$ and one $6s$ bonding electrons at each cerium site and use the method of multipole interaction \cite{NM2}.
According to \cite{Cao,Roo} one $6s$ electron and two $5d$ electrons from
each Ce site contribute to the $(6s \sigma_g)^2 (5d \pi_u)^4$ triple bond. That implies the occupation of
the $Y_{l=2}^{1c}$ and $Y_{2}^{1s}$ orbital $d$-states ($m=\pm1$). The resultant electron density
($(Y_{2}^{1,c}(\Theta,\varphi))^2 + (Y_{2}^{1,s}(\Theta,\varphi))^2$) is axially symmetric.
Expanding it in the multipolar series one finds that
the first nonspherical component responsible for splittings is given by the $Y_{\ell=2}^0$ angular dependence of the $5d$-state density,
\begin{eqnarray}
     \rho_{d,\ell=2}(r,\Theta,\varphi) = 2 C_{d,\ell=2}^0\, R_d^2(r)\, Y_{\ell=2}^0(\Theta,\varphi) ,
\label{A1}
\end{eqnarray}
where $R_d(r)$ is the $5d$ atomic function radial dependence,
and $C_{d,\ell=2}^0=0.090112$ is the angular density matrix element \cite{NM2},
\begin{eqnarray}
     C_{d,\ell=2}^0 = \langle 2,1s | 2,0 | 2,1s \rangle = \langle 2,1c | 2,0 | 2,1c \rangle  \nonumber \\
     = \int_{\Omega} d\Omega Y_{l=2}^{1c}(\Omega)  Y_{\ell=2}^0(\Omega) Y_{l=2}^{1c}(\Omega) .
\label{A2}
\end{eqnarray}
Here $\Omega$ stands for the polar angles $(\Theta,\varphi)$.
Now one can calculate the matrix of the Coulomb interaction of the $5d$ density with the $4f$ electron.
(In principle it will have the $\ell=2$ and $\ell=4$ contributions \cite{NM2}, but the $\ell=4$ term
is small and in the following will be ignored.)
The matrix is diagonal in the space of the $4f$ orbitals with the diagonal elements given by
\begin{eqnarray}
     V_{f,\tau} = V_{df}\, C_{f,\ell=2}^0(\tau) .
\label{A3}
\end{eqnarray}
Here $\tau=(m,c)$ or $(m,s)$ ($m=0-3$) is the azimuthal dependence of the $4f$ orbital functions,
$C_{f,\ell=2}^0(\tau)=\langle 3,\tau | 2,0 | 3,\tau \rangle$ [compare with (\ref{A2})], while
\begin{subequations}
\begin{eqnarray}
     & & V_{df} = 2 C_{d,\ell=2}^0\, v_{2,df} ,  \label{A4a} \\
     & & v_{2,df} = \frac{4\pi}{5} \int \int dr\, dr'\, R_d^2(r) \frac{r_<^2}{r_>^3}  R_f^2(r') . \label{A4b}
\end{eqnarray}
\end{subequations}
$r_<$ ($r_>$) is the largest (the smallest) value of $r$ and $r'$.
$v_{2,df}$ describes the quadrupolar ($\ell=2$) component of the $5d-4f$ intrasite Coulomb repulsion \cite{NM2}.
With the $5d$ and $4f$ atomic radial dependencies (DFT-LDA atomic calculations \cite{DFT}) we obtain
$v_{2,df}=7.1054$ eV and $V_{df}=1.2806$ eV.
The $C_{f,\ell=2}^0(\tau)$ coefficients and the corresponding $4f$ energy splittings are quoted in Table \ref{table5}.
(Notice that due to the $z-$axial symmetry $C_{f,\ell=2}^0(m,c)=C_{f,\ell=2}^0(m,s)$.)
%
\begin{table}
\caption{Coefficients $C_{f,\ell=2}^0(\tau)$ [$\tau$=$(m,c)$ or $(m,s)$] and
calculated splittings $E_{4f}(\tau)$ of the $4f$ states (in eV) caused by the quadupolar Coulomb repulsion with two occupied $5d$ states
($m_d=\pm1$), see text for details.
\label{table5}     }

\begin{ruledtabular}
 \begin{tabular}{ c  c c c c  }
           $\tau$       & $(m=0)$   &  $(1,c)$, $(1,s)$ & $(2,c)$, $(2,s)$ & $(3,c)$, $(3,s)$ \\

 $C_{f,\ell=2}^0(\tau)$ & 0.16821   &  0.12616          & 0                & -0.21026   \\
 $E_{4f}(\tau)$         & 0.21540   &  0.16156          & 0                & -0.26925   \\

 \end{tabular}
\end{ruledtabular}
\end{table}

Thus, we conclude that the $5d-4f$ Coulomb repulsion favors the lowering of the $m=\pm3$ $f-$states.
However, there are two other interactions
which act in the opposite direction: the screened attraction to the other nucleus and the $5d-4f$ exchange interaction.

The other nucleus attraction leads to
an axially symmetric Coulomb potential which has the form of (\ref{A3}), but with an interaction constant of the opposite sign.
The $5d-4f$ exchange interaction is more complex.
In terms of the intrasite Coulomb multipole repulsion it corresponds to the odd ($\ell=1,3,5$) transition
density contributions, and the dipolar ($\ell=1$) contribution is rather strong \cite{NM2}.
However, this exchange interaction leads to the high spin ($S=2$) ground state
(the $^5 H_3$ term of the $4f5d^26s$ configuration) of the cerium atom,
which in the real molecule is effectively suppressed by the triple bond formation with a zero spin state of
the binding $5d$ and $6s$ states.
Therefore, the bond formation mechanism diminishes the $4f-5d$ exchange interaction,
while the $4f-5d$ Coulomb repulsion (\ref{A3}) remains and lowers
the $m=\pm3$ $f-$states in respect to the others \cite{Cao,Roo}.


\section{Appendix B}

The relevant Coulomb integrals, Eq.\ (\ref{f4}), can be simplified by considering
their azimuthal parts, i.e. the azimuthal dependence of the corresponding $\psi_{o_1,a}(\vec{r}_1)\, \psi_{o_3,a}(\vec{r}_1)$ functions
standing on the right and their counterparts $\psi_{o_2,b}(\vec{r}_2)\, \psi_{o_4,b}(\vec{r}_2)$ on the left
from the Coulomb interaction $V(\vec{r}_1,\vec{r}_2)=1/r_{12}$, Eq.\ (\ref{h2}). In the following we define
$\vec{r}_a=\vec{r}_1-\vec{R}_a$ and $\vec{r}_b=\vec{r}_2-\vec{R}_b$.
The important observation is that the Coulomb integral differs from zero only if the azimuthal density dependencies on the left and on the right side
have the same functional form.
This conclusion can be derived by various ways, in particular it follows from the multipole expansion of the Coulomb interaction
\begin{eqnarray}
  V(\vec{r}_a,\vec{r}_b)=
  \sum_{l,\tau}  v_{l}(r_a,r_b)\,
  Y_l^{\tau}(\hat{r}_a)\, Y_l^{\tau}(\hat{r}_b) .
\label{b1}
\end{eqnarray}
Here the angular functions $Y_l^{\tau}(\hat{r}_a)$ and $Y_l^{\tau}(\hat{r}_b)$ have the same azimuthal dependence ($\tau$)
which is a trivial conclusion if they refer to the same site: $\vec{R}_a=\vec{R}_b$.
In case of two different sites (aligned along the $z-$axis)
one can expand a spherical harmonic centered at one site in terms of the spherical
harmonics centered at the other site \cite{NM5}. It turns out that the procedure conserves the
azimuthal dependence of the spherical harmonics translated along the $z-$axis \cite{NM5},
which proves our initial statement.

In the following in deriving relations between the Coulomb integrals, Eq.\ (\ref{f4}),
we use chemist's notations for the matrix elements \cite{Sza}. Further, we explicitly single out the azimuthal dependence
$\rho_{o_1,\, o_3}(\varphi)$ out of the density functions $\psi_{o_1,a}(\vec{r}_a)\, \psi_{o_3,a}(\vec{r}_a)$ with
$o_1,\, o_3=c,\, s$, Eq.\ (\ref{f1a},b).
We then arrive at three different dependencies:
\begin{subequations}
\begin{eqnarray}
   & & \rho_{c,c}(\varphi) = (\cos 3\varphi)^2 = \frac{1}{2} + \frac{1}{2}\cos 6 \varphi , \label{b2a} \\
   & & \rho_{s,s}(\varphi) = (\sin 3\varphi)^2 = \frac{1}{2} - \frac{1}{2}\cos 6 \varphi , \label{b2b} \\
   & & \rho_{c,s}(\varphi) = \cos 3\varphi \, \sin 3\varphi = \frac{1}{2}\sin 6 \varphi  .
\label{b2c}
\end{eqnarray}
\end{subequations}
From the azimuthal selection rule it follows that $(c_a,c_a| V |c_b,s_b)=D=0$, Eq.\ (\ref{f4}) and (\ref{f5b}),
and $(c_a,c_a| V |c_b,c_b)=(s_a,s_a| V |s_b,s_b)=B$.

In order to prove the relation (\ref{f5a}) we consider the structure of the general Coulomb integral,
\begin{eqnarray}
  (1_a,3_a|V|2_b,4_b) = \sum_{M} \int_{\Omega} d \Omega f(\Theta)\, I_A(M_a|M_b) ,
\label{b3}
\end{eqnarray}
where $\Theta$ stands for $(\Theta_a,\Theta_b,r_a,r_b)$,
while $I_{A}(M_a|M_b)$ is the azimuthal integral with two identical $\varphi$-dependencies $M_a$ and $M_b$.
For example, $I_{A}(6_a|6_b)$ with $M_a=M_b=M=6$ stand for the azimuthal integral
between $\cos 6 \varphi$ and $\cos 6 \varphi$ or between $\sin 6 \varphi$ and $\sin 6 \varphi$:
\begin{eqnarray}
  I_{A,\, c}(M_a|M_b) &=& \int_0^{2\pi} d\varphi_a \int_0^{2\pi} d\varphi_b \nonumber \\
  & & \cos (M\varphi_a) V(r_a,r_b) \cos (M\varphi_b) . \quad \quad
\label{b3a}
\end{eqnarray}
Each term on the right hand side of (\ref{b3}) has the following form:
\begin{eqnarray}
  \int_{\Omega} d \Omega f(\Theta)\, I_{A}(M_a|M_b) = \int \sin \Theta_a d\Theta_a \int r_a^2\, dr_a
  \nonumber \\
  \int \sin \Theta_b d\Theta_b \int r_b^2 dr_b \,
  C^2 \, [P_{3}^3( \cos \Theta_a ) P_{3}^3( \cos \Theta_b )]^2 \nonumber \\
  R_f^2(r_a) R_f^2(r_b)\, I_{A}(M_a|M_b). \quad \quad
\label{b4}
\end{eqnarray}
Here $P_{l=3}^{m=3}(x)$ is the associated Legendre polynomial \cite{BC}, $C=-\sqrt{70/\pi}/8$, $R_f(r)$ is the radial dependence of the $4f$ functions.
Taking into account three relevant azimuthal dependencies (\ref{b2a})-(\ref{b2c}) we get
\begin{eqnarray}
   (c_a,c_a|V|s_b,s_b) &=& \frac{1}{4} \int_{\Omega} d \Omega f(\Omega) ( I_{A}(0_a,0_b) - I_{A} (6_a | 6_b)) , \nonumber \\
   (c_a,c_a|V|c_b,c_b) &=& \frac{1}{4} \int_{\Omega} d \Omega f(\Omega) ( I_{A}(0_a,0_b) + I_{A} (6_a | 6_b)) , \nonumber \\
   (c_a,s_a|V|c_b,s_b) &=& \frac{1}{4} \int_{\Omega} d \Omega f(\Omega)\, I_{A}(6_a | 6_b )  .
\label{b5}
\end{eqnarray}
From Eqs.\ (\ref{b5}) we immediately obtain
\begin{eqnarray}
  (c_a,c_a|V|c_b,c_b) - (c_a,c_a|V|s_b,s_b) = 2 (c_a,s_a|V|c_b,s_b), \nonumber \\
\label{b6}
\end{eqnarray}
or $B-A=2C$, Eq.\ (\ref{f5a}).

For the exchange charge $\rho_{ab}(\vec{r})=\psi_{o_1,a}(\vec{r})\, \psi_{o_2,b}(\vec{r})$ ($o_1,o_2=c,s$) which corresponds to the
$4f$ electron transition from the site $a$ to the site $b$ we again have the three azimuthal dependencies, Eqs.\ (\ref{b2a})-(\ref{b2c}).
Therefore, proceeding as before albeit with a modified expressions for the integrals we obtain the analogous relation
\begin{eqnarray}
  (c_a,c_b|V|c_a,c_b) - (c_a,c_b|V|s_a,s_b) = 2 (c_a,s_b|V|c_a,s_b), \nonumber \\
\label{b7}
\end{eqnarray}
or $B'-A'=2C'$, Eq.\ (\ref{f12}). As before, the result is based solely on the azimuthal selection rule.

\section{Appendix C}

The action of the $z-$component of the orbital momentum on the real spherical harmonics $Y_l^{m,c}$, $Y_l^{m,s}$ \cite{BC}
(the $z$-axis coincides with the molecular axis) is given by
\begin{eqnarray}
   L_z\, Y_l^{m,c} = mi Y_l^{m,s}, \quad  L_z\, Y_l^{m,s} = -mi Y_l^{m,c} .
\label{c1}
\end{eqnarray}
The matrix elements of $L_z=L_{1,z}+L_{2,z}$ in the space of two electron $4f$ states (Eqs.\ (\ref{f2a}) and (\ref{f2b})) can be obtained from these relations.
Explicitly, the matrix of $L_z$ has the nonzero blocks $L_z(s,s)$ ($s=1-4$) [see Eq.\ (\ref{f3a})] of the form
\begin{eqnarray}
     L_z(s,s) = \mu_B \begin{pmatrix} 0 & 0 & 3 i & -3 i \\
                                        0 & 0 & 3 i & -3 i \\
                                     -3 i & -3 i & 0 & 0 \\
                                      3 i &  3 i & 0 & 0  \end{pmatrix} ,
\label{c2}
\end{eqnarray}
where $\mu_B$ is the Bohr magneton.

Further, the 16$\times$16 spin matrix $S_z$, Eq.\ (\ref{f3a}), has two nonzero diagonal blocks $S_z(1,1)$ and $S_z(4,4)$ of the form
\begin{eqnarray}
     S_z(s, s) = \pm \mu_B \hat{1} ,
\label{c3}
\end{eqnarray}
where the plus sign refers to the $\alpha,\, \alpha$ block ($s,s=1,1$) and the minus sign to the $\beta,\, \beta$ block ($s,s=4,4$),
while $\hat{1}$ stands for the unit 4$\times$4 matrix.

The molecular interaction with an external magnetic field $H$ is given by
\begin{eqnarray}
     V_{mag} = - \vec{\cal{M}} \vec{H},
\label{c4}
\end{eqnarray}
with the molecular magnetic moment
\begin{eqnarray}
     \vec{ \cal{M} } = 2\vec{S} + \vec{L} .
\label{c5}
\end{eqnarray}
For the magnetic field $H$ applied along the $z-$axis we obtain
\begin{eqnarray}
   V_{mag}^z = - h(2S_z+L_z) ,
\label{c6}
\end{eqnarray}
where $h=H \mu_B$.
The relations (\ref{c2}), (\ref{c3}) and (\ref{c6}) are used to calculate the $z$-component of the orbital, spin
and full momentum for each two-electron level.

For example, by means of Eq.\ (\ref{c2}) we can find the orbital quantum number required for
the characterization of the degenerate states, which are considered in Sec.\ \ref{subsecA}.
There we have discussed the three energy levels given by Eq.\ (\ref{f6a})-(\ref{f6c}).
The corresponding eigenvectors are
\begin{subequations}
\begin{eqnarray}
 & & (-1,1,0,0)/\sqrt{2},   \label{c7a} \\
 & & (1,1,0,0)/\sqrt{2}, \quad (0,0,-1,1)/\sqrt{2},  \label{c7b} \\
 & & (0,0,1,1)/\sqrt{2}. \label{c7c}
\end{eqnarray}
\end{subequations}

In the presence of the external magnetic field the Hamiltonian $V(f-f)$ of the direct Coulomb repulsion (see Sec.\ \ref{subsecA})
is modified:
\begin{eqnarray}
   H = E_{bond} + V(f-f) + H_L ,
\label{c7}
\end{eqnarray}
where the last term is the orbital polarization
given by $H_L=-L_z h$ (the first term $E_{bond}$ is the binding energy of Ce$_2$ which is not important for our discussion).
From (\ref{c7}) we obtain four eigenvalues ($i=1-4$)
\begin{eqnarray}
  E_i-E_0 = \quad -C,\quad C-6h, \quad C+6h, \quad 3C ,
\label{c8}
\end{eqnarray}
and the corresponding eigenvectors:
$(-1,1,0,0)/\sqrt{2}$, $(i,i,-1,1)/\sqrt{2}$, $(i,i,1,-1)/\sqrt{2}$, and $(0,0,1,1)/\sqrt{2}$.

Notice, that at zero magnetic field ($h \rightarrow 0$) $E_2=E_3$ as in (\ref{f6b}).
On the other hand, from (\ref{c8}) it is clear that
two components correspond to $M_z=\pm6$ and therefore in this doublet $L_z=6$.
In other two non-degenerate states $L_z=M_z=0$ (i.e. they are $\Sigma$-states).



\end{document}